\documentclass[fleqn,usenatbib]{mnras}

\usepackage{newtxtext,newtxmath}

\usepackage[T1]{fontenc}

\DeclareRobustCommand{\VAN}[3]{#2}
\let\VANthebibliography\thebibliography
\def\thebibliography{\DeclareRobustCommand{\VAN}[3]{##3}\VANthebibliography}

\usepackage{graphicx}	
\usepackage{amsmath}	

\title[Circumbinary Matter in Hot Subdwarfs]{Detection of Ubiquitous Circumbinary Matter in Hot Subdwarfs Formed from Common-Envelope Ejections}

\author[Jiangdan Li et al.]{
Jiangdan Li,$^{1,2}$ \thanks{E-mail: lijiangdan@ynao.ac.cn}
Christian Wolf,$^{3,4}$
Jiao Li,$^{1}$
Yangping Luo, $^{5}$
Jingkun Zhao, $^{6}$
Bingqiu Chen,$^{7}$
\newauthor Lin Zhang, $^{7}$
Shi Jia,$^{1}$
Xuefei Chen, $^{1,2,8,9}$ 
and Zhanwen Han$^{1,2,8,9}$\thanks{E-mail: zhanwenhan@ynao.ac.cn}
\\
$^{1}$Yunnan Observatories, Chinese Academy of Sciences (CAS), 396 Yangfangwang, Guandu District, Kunming 650216, P.R. China\\
$^{2}$Key Laboratory for the Structure and Evolution of Celestial Objects, CAS, Kunming 650216, P.R. China\\
$^{3}$Research School of Astronomy and Astrophysics, Australian National University, Weston Creek ACT 2611, Australia\\
$^{4}$Centre for Gravitational Astrophysics, Australian National University, Canberra ACT 2600, Australia\\
$^{5}$Department of Astronomy, China West Normal University, Nanchong 637002, P.R. China\\
$^{6}$CAS Key Laboratory of Optical Astronomy, National Astronomical Observatories, Chinese Academy of Sciences, Beijing 100101, P.R. China\\
$^{7}$South-Western Institute for Astronomy Research, Yunnan University, Kunming 650504, China\\
$^{8}$International Centre of Supernovae, Yunnan Key Laboratory, Kunming 650216, P.R. China\\
$^{9}$University of Chinese Academy of Sciences, Beijing 100049, China\\
}
\date{Accepted XXX. Received YYY; in original form ZZZ}

\pubyear{\the\year{}}

\begin{document}
\label{firstpage}
\pagerange{\pageref{firstpage}--\pageref{lastpage}}
\maketitle

\begin{abstract}
The formation of compact binary systems is largely driven by their evolution through a common envelope (CE) phase, crucial for understanding phenomena such as type Ia supernovae and black hole mergers. Despite their importance, direct observational evidence for CE material has been elusive due to the transient nature of these envelopes. Numerical simulations suggest that some envelope material may persist post-ejection. 
In this study,  we investigate circumstellar material (CSM) surrounding hot subdwarf (sdB) stars, focusing on material ejected during the CE phase of binary evolution. We analyze Ca II K absorption lines in 727 sdB candidates from the LAMOST-LRS survey, selecting 145 stars with strong absorption features, indicating the presence of CSM. We compare the velocities of the Ca II K lines with the systemic velocities of sdB binaries, confirming that the material originates from ejected common-envelope material. The results show that the CSM persists long after the CE event, suggesting the formation of a stable, long-lived circumstellar environment around sdB stars. This study enhances our understanding of the role of CSM in post-CE evolution and provides new insights into the physical processes shaping the evolution of sdB binaries.

\end{abstract}




\section{Introduction}

Compact binaries are fundamental to modern astrophysics, comprising diverse systems such as double black holes, neutron stars, white dwarfs, X-ray binaries, cataclysmic variables, progenitors of type Ia supernovae, and hot subdwarf binaries. Their formation is predominantly driven by the common envelope (CE) evolution process, first proposed by \cite{Webbink1975PhDT} and \cite{ Paczynski1976IAUS_CE} nearly half a century ago. This process occurs when one star in a binary system expands to fill its Roche lobe, engulfing its companion and forming a common envelope around both stars.

CE evolution stands as the most crucial process in binary evolution, yet it remains one of the least understood aspects \citep{Ivanova2020ceebook}. Although significant progress has been made in numerical simulations, our understanding is still incomplete (i.e. \cite{Passy2012ApJ_CE, Ivanova2016MNRAS, Clayton2017, Iaconi2018MNRAS_CE, Kramer2020A&A_sdB_CEsimulations, Sand2020A&A_CE, Glanz2021MNRAS_CE, Gonzalez-Bolivar2022MNRAS_CE}). The inherent complexity of the underlying physics, along with the need for high spatial resolution and short time steps in simulations, poses considerable challenges. Additionally, the rapid and dynamic nature of the CE interaction complicates modeling efforts during this phase \citep{Ivanova2013AAPR, Ropke2023_CEsimulations}.

While observations offer the potential to validate CE theory, the dynamic and short-lived nature of CE events makes detection challenging \citep{Ivanova_Natalia_Pod2003, Ivanova2013AAPR}. Direct observational evidence has been scarce for decades, despite indications that extremely short-orbital period binaries likely originate from CE events. Two main challenges hinder our understanding: the short timescale of CE events, typically between 1\,000 and 10\,000 years, and the similarity in observational properties between these binaries and normal stars during the CE phase \citep{Chenxuefei2024_binary_review}. 

At the termination of CE evolution, significant material is expected to be ejected, even if the process ultimately leads to the merger of the binary system \citep{Hettinger2015ApJ_merger}. \cite{Ivanova2013Sci} suggested that the radiation from this ejected matter is primarily governed by a recombination front during its cooling phase, which establishes a stable photosphere and generates a plateau shape in the light curves as the material expands. This modeling of plateau luminosity and timescale is illustrated in Figure 1 of \cite{Ivanova2013Sci}, particularly in relation to luminous red novae (LRNe) like V1309 Sco, M85 OT, M31-RV, and V838 Mon—events characterized by unique light curves believed to arise from CE ejections. The brightness of LRNe is attributed to the recombination energy released as the ejected envelope expands, with a plateau phase indicating a recombination front \citep{Ivanova2013Sci}. An effective photospheric temperature of approximately 5\,000 K for thick ejecta results in red outburst events, and once the envelope has fully recombined, it may suddenly become transparent. These characteristics, along with ejection velocities and event rates, align closely with observational data for LRNe.

Following the pioneering work of \cite{Ivanova2013Sci}, interest in LRNe has surged. Notable examples include two LRNe in nearby galaxies (M101 OT2015-1 and M31LRN 2015) that exhibited pre-outburst sources \citep{Blagorodnova2017ApJ, MacLeod2017ApJ}. V1309 Sco, which resulted from the merger of a contact binary system, serves as a significant case. Extensive monitoring through the OGLE survey allowed for a detailed reconstruction of its pre-outburst evolution, revealing a contact binary with an orbital period of about 1.4 days, which decreased until the merger in March 2008 \citep{Tylenda2011A&A}. This provided direct evidence that contact binaries can culminate in mergers, producing eruptions similar to V838 Mon-type events. Subsequent observations of V1309 Sco have uncovered a complex circumstellar environment. The molecular gas remains relatively cool (35 to 113 K), while atomic gas reaches higher temperatures (5 to 15 kK) due to shock heating \citep{Steinmetz2024A&A_V1309}. Kinematic studies suggest an asymmetric bipolar structure, a feature observed in other Galactic LRNe, indicating the significant role of bipolar outflows in shaping post-merger environments.

Some peculiar single stars, including magnetic stars, blue stragglers, rapid rotators, and T CrB stars, are thought to have formed from the merging of binaries through CE evolution \citep{Han2020}. An intriguing example is TYC 2597-731-1, which features an unusual ring-shaped ultraviolet nebula surrounding it. This object is likely in an evolutionary stage between the dynamic onset of CE ejection and its theorized final equilibrium state \citep{Hoadley2020Natur}. Despite its classification as an old star, TYC 2597-731-1 exhibits abnormally low surface gravity and a detectable long-term luminosity decay, which is uncharacteristic for its evolutionary stage. Observations suggest that TYC 2597-731-1 merged with a lower-mass companion several thousand years ago, offering direct insights into the merging process of two stars into a single entity \citep{Hoadley2020Natur}.

Recent high-resolution observations by ALMA have identified evolved stars with high-velocity water maser jets, indicating significant mass loss rates associated with CE ejections \citep{Khouri2022NatAs}. These findings underscore the importance of studying CE processes within the broader context of binary evolution, as they provide critical insights into the physical mechanisms governing mass transfer and stellar interactions.

In particular, ejected CE material has been detected around the short-period subdwarf O (sdO) + white dwarf (WD)  binary J1920-2001 \citep{Lijiangdan2022MNRAS}. This binary system has an orbital period of 3.5 hours and a mass ratio of 0.738, with the sdO star having a mass of 0.55 M$_{\odot}$ and overflowing its Roche lobe, likely transferring mass to the WD through an accretion disk. According to models for hot subdwarf (sdB) stars \citep{Han2002, Han2003}, J1920-2001 likely formed through a CE ejection channel, where the progenitor of the sdO star was either a red giant branch (RGB) star or an early asymptotic giant branch (AGB) star. Notably, strong Ca H\&K lines in this system have been observed to be blue-shifted by approximately 200\,km/s, suggesting the presence of CE material ejected around 10\,000 years ago \citep{Hoadley2020Natur}. The detection of ejected CE around short-period sdB binaries is not surprising, as these systems primarily form through CE ejections. With their relatively short lifetimes of about 10 million years, hot subdwarf binaries like J1920-2001 present a unique opportunity to investigate the remnants of CE ejections and their role in binary evolution. By applying similar methods to the LAMOST survey, we anticipate discovering more sdB binary candidates, which will significantly enhance our understanding of the CE process and inform current binary evolution theories.

Our study centers on hot subdwarf binaries, a unique class of post-CE systems that represent earlier evolutionary stages compared to post-AGB and post-CE binaries found in planetary nebulae (PNe) and LRNe. 
In this article, we investigate the detection of circumbinary matter formed through common envelope ejections in hot subdwarf binaries. By providing direct empirical evidence of these phenomena, our study aims to deepen the understanding of common envelope evolution and its impact on binary star formation and evolution. The paper is structured as follows: Section \ref{sec:sample} describes the sdB sample selection from the LAMOST-LRS database. Section \ref{sec:methods} outlines the analytical approaches, including spectral analysis of Ca II K lines and techniques for identifying circumbinary matter. Section \ref{sec:results} presents the observational results, discussing circumstellar medium (CSM) densities and detailed parameters for 727 sdBs. Section \ref{sec:ubiquity} explores the ubiquity of circumbinary material, incorporating evolutionary tracks for sdBs, RV analyses, RV curves for select stars, and the distribution of Ca II K line RVs relative to spectral data. Section \ref{sec: Discussions} examines key topics, including the mass and radius of circumbinary material, bias correction for ISM blending, and insights from Na I D and K I line analyses. Finally, Section \ref{sec: Conclusion} summarizes our findings and highlights future directions for research, emphasizing opportunities to further investigate this crucial aspect of astrophysics.

\section{Sample Selection}\label{sec:sample}

The data for this study were obtained from the LAMOST-LRS DR7 survey, utilizing the catalog of hot subdwarf star candidates published by \cite{Luo2021}. This catalog is derived from the Gaia DR2 catalog of hot subdwarf star candidates \citep{Geier2019}. After rejecting objects with poor spectra, main-sequence stars, white dwarfs, and those with strong Mg I (5183 {\AA}) or Ca II (8650 {\AA}) absorption lines, \cite{Luo2021} identified 1\,587 sdB candidates with spectral signal-to-noise ratios (SNRs) greater than 10 in the $g$-band. These candidates were visually matched with reference spectra of hot subdwarf stars, which significantly reduces the possibility of including composite binaries, such as sdB+FGK systems. All the sdBs in the sample are single-lined spectroscopic binaries, likely to be either single sdBs or sdB+white dwarf (WD) or sdB+dM systems.
Within the Gaia DR2 catalog, it is estimated that over 95\% of these candidates are sdB stars, while the remainder are likely white dwarfs or main-sequence stars of spectral types B and O \citep{Geier2019}. The selection criteria established by \cite{Luo2021} ensure a high fidelity in identifying sdB stars, which is further supported by our use of their catalog for our analysis.

LAMOST is a quasi-meridian reflecting Schmidt telescope located at the Xinglong Station of the National Astronomical Observatory in China \citep{Cuixiangqun2012RAA, Zhaogang2012RAA}. It is equipped with 4\,000 optical fibers across a 5-degree field of view (FoV), enabling simultaneous spectra acquisition from 4\,000 celestial objects \citep{Cuixiangqun2012RAA, Zhaogang2012RAA}. The LAMOST-LRS spectra, with a spectral resolution of approximately 1\,800, cover a wavelength range from 3\,800\,{\AA} to 9\,100\,{\AA} and feature prominent spectral lines of H, He, Fe, Mg, and Ca \citep{Cuixiangqun2012RAA}. The spectra are heliocentrically corrected.

To determine the atmospheric parameters and radial velocities of the sdB candidates, \cite{Luo2021} employed TLUSTY synthetic spectra \citep{Hubeny&Lanz1995} as templates. They calculated the effective temperature ($T_{\rm eff}$), surface gravity ($\log{g}$), helium abundance ($\log{y}$), and radial velocity ($RV$). Notably, the Ca II absorption line is absent in the spectra of OB stars due to their high effective temperatures, which ionize most calcium atoms to Ca III \citep{Gray2009book}. 
Consequently, any Ca II absorption along the line of sight must arise from cooler intervening material, such as the interstellar medium (ISM) or circumstellar medium (CSM). Similarly, due to the high temperatures of sdB stars, Ca II K absorption lines are also not expected in their spectra.

For a comprehensive analysis of the Ca II K absorption lines at $3934.77$ {\AA} (all subsequent wavelengths are also in vacuum), and the potential presence of ISM or CSM surrounding the sdB stars, we selected a subsample of sdB stars with higher data quality, specifically those with a $g$-band SNR exceeding 30. The typical uncertainties for effective temperature, surface gravity, radial velocity, and distance in this subsample are approximately $\sim$230\,K, $\sim$0.04\,dex, $\sim$25\,km/s, and $\sim$100\,pc, respectively. The radial velocity error of 25\,km/s is influenced by the spectral resolution. 
Additionally, to minimize the impact of He I ($3937.0$ {\AA}), we excluded He-rich sdBs with $\log{y} = \log(n{\rm He}/n{\rm H}) > 0$. The final sample used in this study includes 727 sdB stars. All stars in the sample exhibit effective temperatures exceeding approximately 20\,000\,K. 


\section{Methods}\label{sec:methods}

In this section, we outline the analytical approach used to investigate the properties of the Ca II K absorption lines in the spectra of selected sdB stars, focusing on their implications for the circumstellar medium. Metals play a crucial role in determining the chemistry, ionization state, and temperature of the gas \citep{Sembach2000ApJ_ISM,Murga2015}, and their properties can be probed through absorption line spectroscopy. The spectral range of LAMOST-LRS encompasses several Ca II transitions, including the H and K lines, the red triplet, and quadrupole transitions, which provide valuable insights into the conditions of the circumstellar medium. However, due to significant blending issues with these other lines, our analysis concentrates exclusively on the Ca II K line.

\begin{figure}
    \centering
    \includegraphics[width=0.5\textwidth]{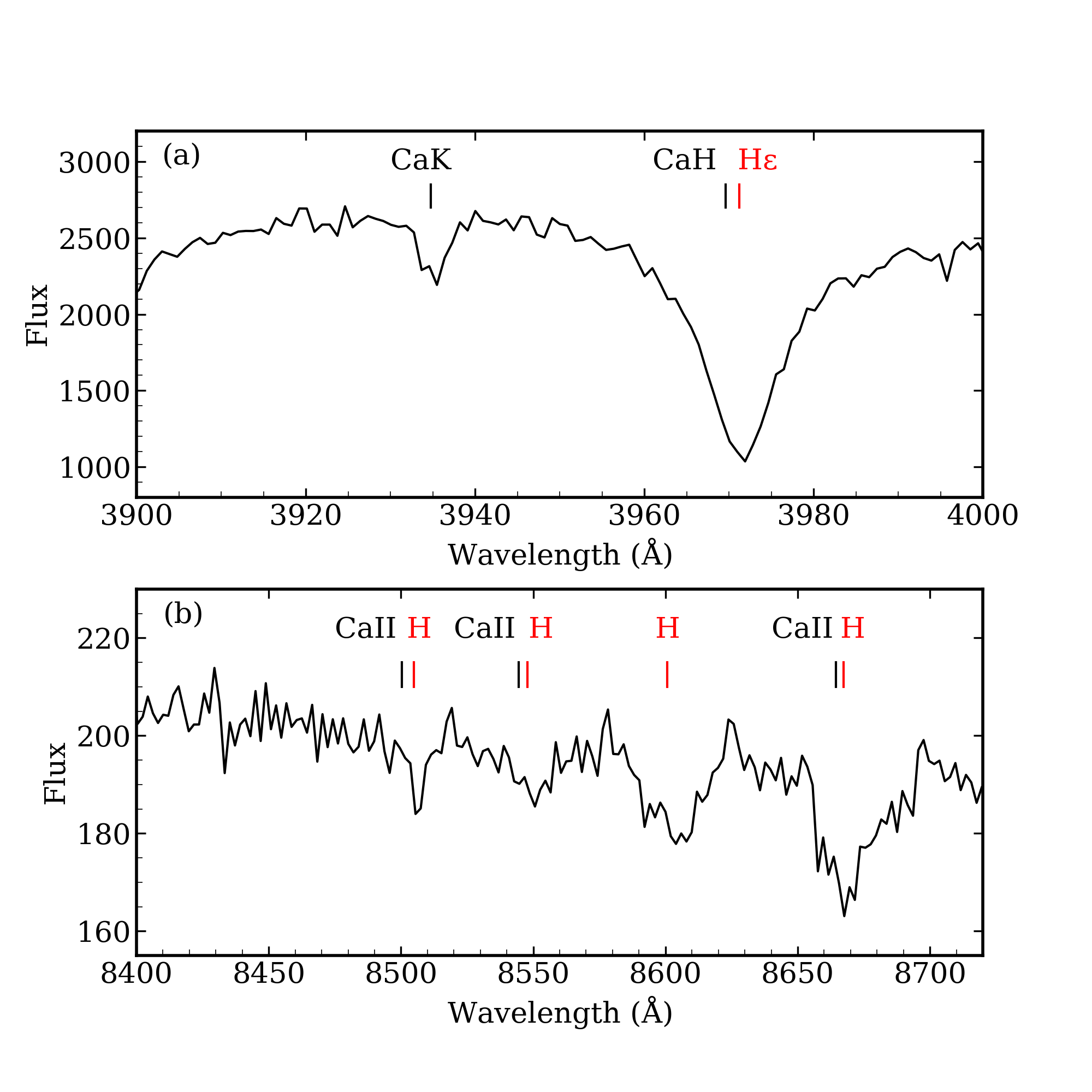}
    \caption
    {We present a randomly selected sdB spectrum from LAMOST-LRS, designated J073228.7+330237.5 in the J2000 frame, with a $g$-band SNR of approximately 71.46. Panel (a) presents the wavelength range 3\,900–4\,000 {\AA}, featuring the Ca II H \& K lines (3\,969.59 and 3\,934.77 {\AA}) along with the strong H$\varepsilon$ line at 3\,971.20 {\AA}, where the Ca II H line is blended with H$\varepsilon$ due to the spectral resolution ($\sim$1800). Panel (b) displays the range 8\,400–8\,720 {\AA}, including the Ca II red triplet lines (8\,500.35, 8\,544.44, and 8\,664.52 {\AA}), closely spaced with the H Paschen series lines (8\,504.83, 8\,547.73, 8\,600.75, and 8\,667.40 {\AA}).
    }
    \label{fig:work3_spectrum_CaII}
\end{figure}

The Ca II H line at 3969.59 {\AA} in vacuum lies very close to the H$\varepsilon$ line at 3971.20 {\AA}, which is particularly strong in sdB star spectra (see Panel (a) of Figure~\ref{fig:work3_spectrum_CaII}). With the LAMOST-LRS resolution of $R \sim 1\,800$, this proximity leads to substantial overlap, making it impossible to fully distinguish the two lines. As a result, interference from H$\varepsilon$ makes it challenging to extract reliable information from the Ca~II~H line.

Similarly, the Ca II red triplet (8500.35 {\AA}, 8544.44 {\AA}, and 8664.52 {\AA}) overlaps with prominent H Paschen lines (8504.83 {\AA}, 8547.73 {\AA}, 8600.75 {\AA}, and 8667.40 {\AA}), which are also strong in sdB spectra (see Panel (b) of Figure~\ref{fig:work3_spectrum_CaII}). The low spectral resolution of LAMOST-LRS further hinders reliable separation of the triplet from the nearby H Paschen lines. Additionally, the quadrupole transitions of Ca II at 7293.48 and 7325.91 {\AA} are weak, and the low SNR in the red band of the spectra makes it difficult to derive effective information.

Including the Ca II H line, red triplet, and quadrupole transitions in our analysis could introduce uncertainties due to blending and would not enhance the results. Therefore, we concentrate on the Ca II K line, which provides the clearest and most reliable data for studying the circumstellar medium around sdB stars.

\subsection{Analysis of Ca~II~K lines}

We analyzed the characteristics of the Ca~II~K line, specifically focusing on its equivalent width ($EW_{\rm CaK}$) and radial velocity ($RV_{\rm CaK}$). To extract these parameters, we employed a single Gaussian fitting function using the Python package {\it SciPy} \citep{2020SciPy-NMeth}. The Gaussian model is expressed as:
\begin{eqnarray}
y=\frac{A}{\sigma \sqrt{2\pi}}e^{-\frac{(x-\mu)^2}{2\sigma^2}}+kx+b,
\end{eqnarray} 
where $A$ represents the depth of the Gaussian curve, $\mu$ is the position of the peak's center, $\sigma$ controls the width of the curve which is the standard deviation, and $k$ and $b$ determine the gradient and the base height of the curve to fit the continuum of an observational spectrum.
The Gaussian function is fitted to the absorption features centered around the spectral wavelength of the Ca~II~K line at 3\,934.77\,{\AA}. 
The position of the Gaussian fit corresponds to the location of the absorption line, and the radial velocity is determined from the wavelength shift of the absorption feature relative to the rest wavelength. The equivalent width is calculated by integrating the area under the Gaussian curve, which corresponds to the absorption feature relative to the surrounding continuum. To account for the continuum, we normalize the absorption by dividing the area by the continuum value ($k\mu + b$). The final $EW$ is derived by averaging 1\,000 random samples drawn from a multivariate normal distribution using the fitted parameters and their respective errors. The uncertainty in $EW$ is given by the standard deviation of these samples.

\begin{figure}
    \centering
    \includegraphics[width=0.5\textwidth]{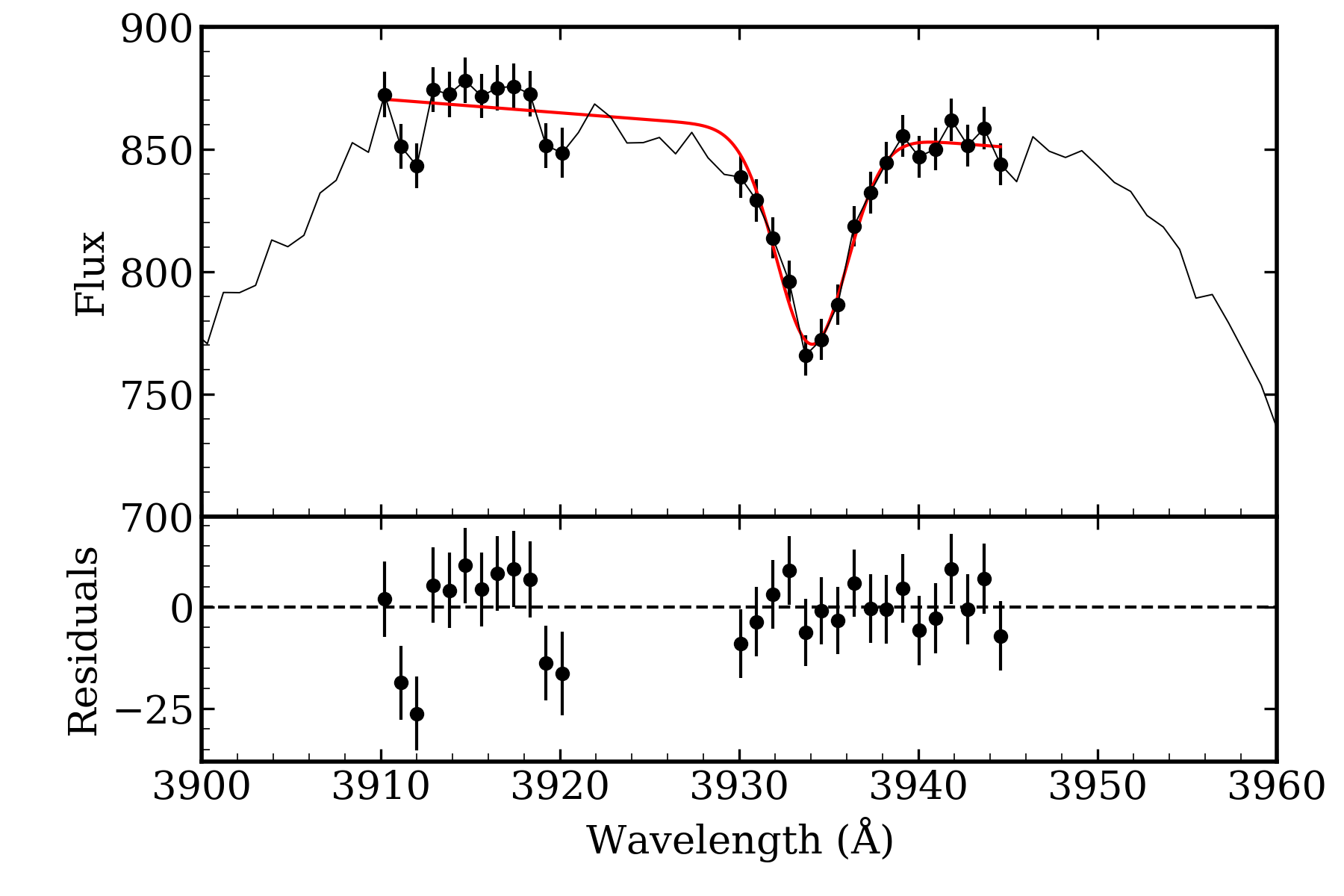}
    \caption
    {
    An example of Gaussian fitting applied to the spectrum of an sdB star from LAMOST-LRS. The upper panel displays the observed spectrum over the wavelength range of 3\,900–3\,960\,{\AA}. Black dots mark the analysis range for the Ca~II~K spectral line, including the line core, its wings, and the surrounding continuum. The red curve represents the Gaussian fit to the Ca~II~K line. The lower panel shows the residuals, illustrating the differences between the observed spectrum and the Gaussian fit.
    }
    \label{fig:work3_example}
\end{figure}

An example of a Gaussian fit applied to the Ca~II~K absorption line is shown in Figure~\ref{fig:work3_example}. In this figure, the central position of the Gaussian fit corresponds to the wavelength of the Ca~II~K absorption line, while the contour highlights the equivalent width of the feature. To define the analysis range for the Ca~II~K spectral line, we adopt a specific wavelength interval based on \cite{Liu2015RAA}, using 3\,910–3\,923\,{\AA} for the left continuum and 3\,930–3\,946\,{\AA} for the line and right continuum.

To ensure the reliability of our results, we conducted a visual inspection of the LAMOST-LRS spectra across the entire sample, selecting only those spectra that clearly exhibited Ca II K (3\,934.77 {\AA}) spectral line features. Consequently, our final analysis included 623 out of the initial 727 hot subdwarf candidates, all of which displayed observable Ca II K absorption lines and reliable measurements for equivalent width and radial velocity.

The absence of observable Ca II K lines in the remaining 104 sdB candidates can be attributed to several factors. First, there may be a lack of Ca II K material or ISM in the region. Additionally, some stars might have lower mass loss rates or insufficient circumbinary material to generate detectable absorption features. Variations in the circumstellar environment can influence the strength of the spectral lines; if the material is dispersed or not concentrated along our line of sight, this could lead to undetectable Ca II K absorption. Certain stars may also possess characteristics that hinder the formation of strong Ca II K lines, such as lower effective temperatures or differing atmospheric compositions, which can affect the ionization balance and the resulting line strengths. Finally, the quality of the LAMOST-LRS spectra may contribute; inadequate spectral resolution or high noise levels could obscure weaker Ca II K features. Thus, a combination of these factors likely accounts for the lack of detectable Ca II K absorption in the 104 sdB candidates, underscoring the significance of the 623 stars identified in our analysis.

\subsection{Correlation Between Ca II K Line Equivalent Width and Reddening}

As previously mentioned, the high surface temperatures of sdB stars result in spectra dominated by H and He absorption lines \citep{Gray2009book}. This implies that any observed Ca II K absorption at 3934.77\,{\AA} must originate either from the ISM or the CSM. To distinguish between these two sources, we first estimated the reddening extinction coefficient $E(B-V)$ for the sdB stars, using it as a proxy for the line-of-sight dust column density. The $E(B-V)$ values were derived using the Bayestar19 3D dust maps \citep{Green2019}, incorporating right ascension (RA), declination (Dec), and distance data from Gaia eDR3 \citep{Luo2021, Gaia_edr3_2021}. These values provide quantitative measures of dust along the line of sight, with higher values indicating denser dust and greater light attenuation \citep{Schlafly2011ApJ_dust, Green2019}.

\begin{figure}
    \centering
    \setlength{\leftskip}{0pt}
    \includegraphics[width=0.5\textwidth]{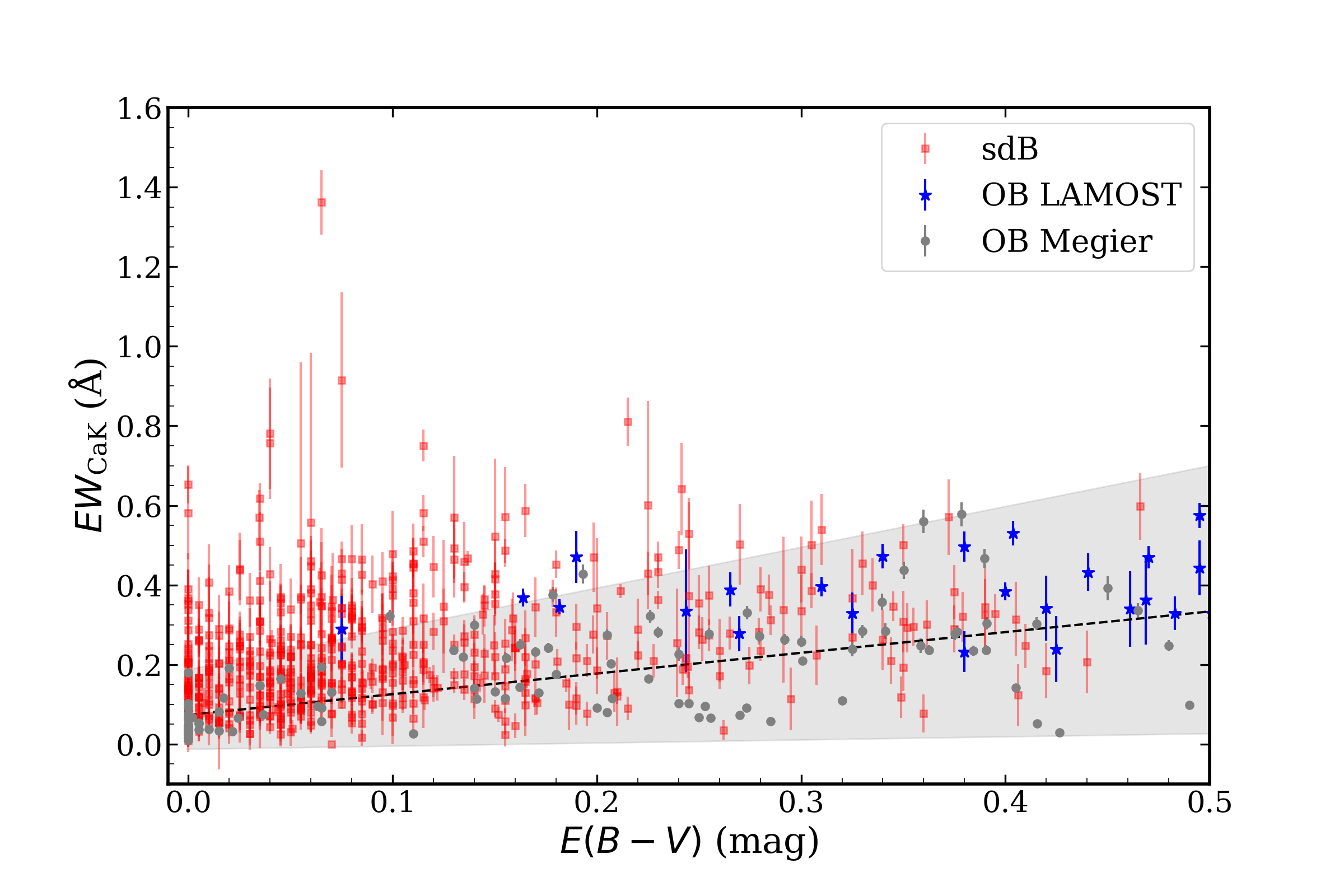}
    \caption
{The correlation between the equivalent width ($EW_{\rm CaK}$) of the Ca~II~K line and the reddening extinction coefficient $E(B - V)$. Gray symbols represent OB stars from \citet{Megier2009A&A}, with the dashed line indicating a linear fit that reflects the typical ISM relationship between $EW_{\rm CaK}$ and $E(B - V)$. The shaded region shows the 95\% confidence interval around this fit. Red squares represent sdB stars observed by LAMOST-LRS, with error bars reflecting the $EW_{\rm CaK}$ measurement uncertainties. Blue stars denote a control sample of OB stars from LAMOST-LRS, with 87\% falling within the shaded ISM region.
}
    \label{fig:work3_disk_EW}
\end{figure}

To establish a baseline ISM relationship, we modeled the dependence of the Ca II K line equivalent width ($EW_{\rm CaK}$) on $E(B-V)$ using a sample of 284 OB field stars from \citet{Megier2009A&A}. For $E(B-V)$ values below 0.5, which are typical for sdB stars, the relationship between $EW_{\rm CaK}$ and $E(B-V)$ remains linear. However, at higher extinction values, the Ca II K line begins to saturate, leading to a flattening of the $EW_{\rm CaK}$ vs. $E(B-V)$ relationship \citep{Munari&Zwitter1997A&A, Smoker2003MNRAS_ISM_cak, Smoker2015A&ACaK, Murga2015}. We fitted a linear relationship between $EW_{\rm CaK}$ and $E(B-V)$ for $E(B-V)<0.5$, using $E(B-V)$ values derived from the Bayestar19 dust maps, consistent with the approach applied to our sdB sample. The resulting best-fit relationship is shown in Figure~\ref{fig:work3_disk_EW}:
\begin{equation}
    EW_{\rm CaK} = 0.52 (\pm 0.06) \times E(B-V) + 0.07 (\pm 0.01),
\end{equation}
where $EW_{\rm CaK}$ is measured in {\AA} and $E(B-V)$ in magnitudes. The dashed line in the figure represents the ISM relationship between $EW_{\rm CaK}$ and $E(B-V)$, while the shaded region shows the 95\% confidence range around this fit.

To validate the ISM relationship, we created a control sample of OB stars from the LAMOST-LRS survey \citep{XiangMaosheng2022A&A_LAMOSTOB}. We selected OB stars with effective temperatures above 20\,000 K, surface gravities in the range $3.5 < \log{g} < 5$, and absolute G-band magnitudes below 2. To avoid contamination from circumstellar effects, we excluded OB stars with H$\alpha$ emission lines (which indicate stellar winds or circumstellar material), as well as those with Mg II absorption at 4\,482.38 {\AA} stronger than He I 4\,471.48 {\AA}, characteristic of cooler A-type stars \citep{Gray2009book}, and those with strong He I ($3\,937.0$ {\AA}), which can influence the $EW$ of Ca II K lines. The sample was further restricted to stars within a distance of 5\,000 pc, consistent with the range of sdB stars. This yielded a final sample of 23 OB stars with $E(B-V)$ values smaller than 0.5 (shown as the blue stars Figure~\ref{fig:work3_disk_EW}).

We then analyzed the Ca II K lines using Gaussian fitting to determine the equivalent widths, following the same procedure applied to our sdB stars. The OB star sample closely followed the ISM relationship established by \citet{Megier2009A&A}, confirming the alignment with ISM expectations. Among the OB stars with $E(B-V) < 0.5$, the median $EW_{\rm CaK}$ for stars without H$\alpha$ emission was 0.37 {\AA}, while for stars with H$\alpha$ emission, the median $EW_{\rm CaK}$ increased to 0.43 {\AA}, suggesting that circumstellar material enhances Ca II K absorption. A total of 20 OB stars fell within the shaded region around the linear fit, yielding an 87\% confidence level that the fit accurately represents the ISM relationship between $EW_{\rm CaK}$ and $E(B-V)$.

Given that intrinsic absorption of Ca II K line is not expected in atmospheres of sdBs \citep{Gray2009book}, we anticipate that observed $EW_{\rm CaK}$ values of should conform to this ISM extinction relationship. Deviations from this trend—where the observed $EW_{\rm CaK}$ significantly exceeds predictions based on $E(B-V)$ values from dust maps \citep{Green2019}—suggest the presence of additional circumstellar material. In these cases, the hot subdwarfs are likely in binary systems with ejected common envelopes contributing to the observed absorption. Our analysis identified 188 sdB stars with Ca II K absorption exceeding ISM-only predictions, suggesting that these stars are likely surrounded by remnants of common envelopes. This selection process improves the reliability of our sample, highlighting stars with significant excess Ca II K absorption likely indicative of circumstellar material, and offering valuable insights into binary interactions and sdB evolutionary pathways.

\subsection{Comparison of Local and Large-Scale Reddening Measurements}

Comparing $E(B-V)$ values derived from large-scale dust maps and those from spectral energy distribution (SED) fitting offers insights into the presence of circumstellar material around stars. Discrepancies between these values suggest higher densities of circumstellar material compared to the surrounding interstellar medium, particularly in regions not fully captured by broad, line-of-sight dust maps.

For this analysis, we used the Bayestar19 3D dust maps, which provide averaged dust reddening values across the line of sight, inferred probabilistically from approximately 4.21 million spatial pixels \citep{Green2019}. While these maps give a broad estimate of dust, they may not fully capture dense, localized structures around the star. In contrast, $E(B-V)$ values derived from multi-wavelength photometry through SED fitting offer a more localized measure of extinction affecting each star, capturing circumstellar and nearby interstellar dust more accurately.
By comparing the SED-derived $E(B-V)$ values with those from Bayestar19, we can identify stars where the local reddening differs significantly from the broader estimate, indicating the presence of excess local dust and gas. These stars, likely surrounded by circumstellar material, are the focus of our further analysis.

To estimate the reddening coefficient $E(B-V)$ for our sample of sdB stars, we performed SED fitting using data from GALEX \citep{GALEX2007}, Gaia EDR3 \citep{Gaia_edr3_2021}, APASS \citep{Henden2014CoSka_APASS}, 2MASS \citep{Skrutskie2006AJ_2MASS}, and WISE \citep{WISE2010}. Using the T{\"u}bingen non-local thermodynamic equilibrium (NLTE) Model-Atmosphere package (TMAP) \citep{Werner2003} and the SPEEDYFIT package \citep{Vos2012, Vos2013, Vos2017}, we employed a Markov Chain Monte Carlo (MCMC) approach to minimize residuals between observed and theoretical SEDs, incorporating prior constraints on effective temperature, surface gravity, and distance \citep{Luo2021, Gaia_edr3_2021}. In this analysis, we assume that the sdB star contributes 100\% of the observed light, both in the SED fitting and in the region around the Ca II K line. This assumption is justified by the rigorous sample selection process described in \cite{Luo2021}, which excluded objects displaying features indicative of composite binaries, such as Mg I (5183 {\AA}) and Ca II (8650 {\AA}) absorption lines. Furthermore, the absence of any infrared excess in our SED fitting strongly supports the conclusion that our sample contains very few, if any, composite binaries. Consequently, the Ca II K line is neither diluted by flux from a companion star nor affected by contamination from composite systems, confirming its origin as interstellar or circumstellar.
This process yielded an $E(B-V)$ value specific to each sdB star, which we used to analyze the circumstellar material properties.

\begin{figure}
    \centering
    \includegraphics[width=0.5\textwidth]{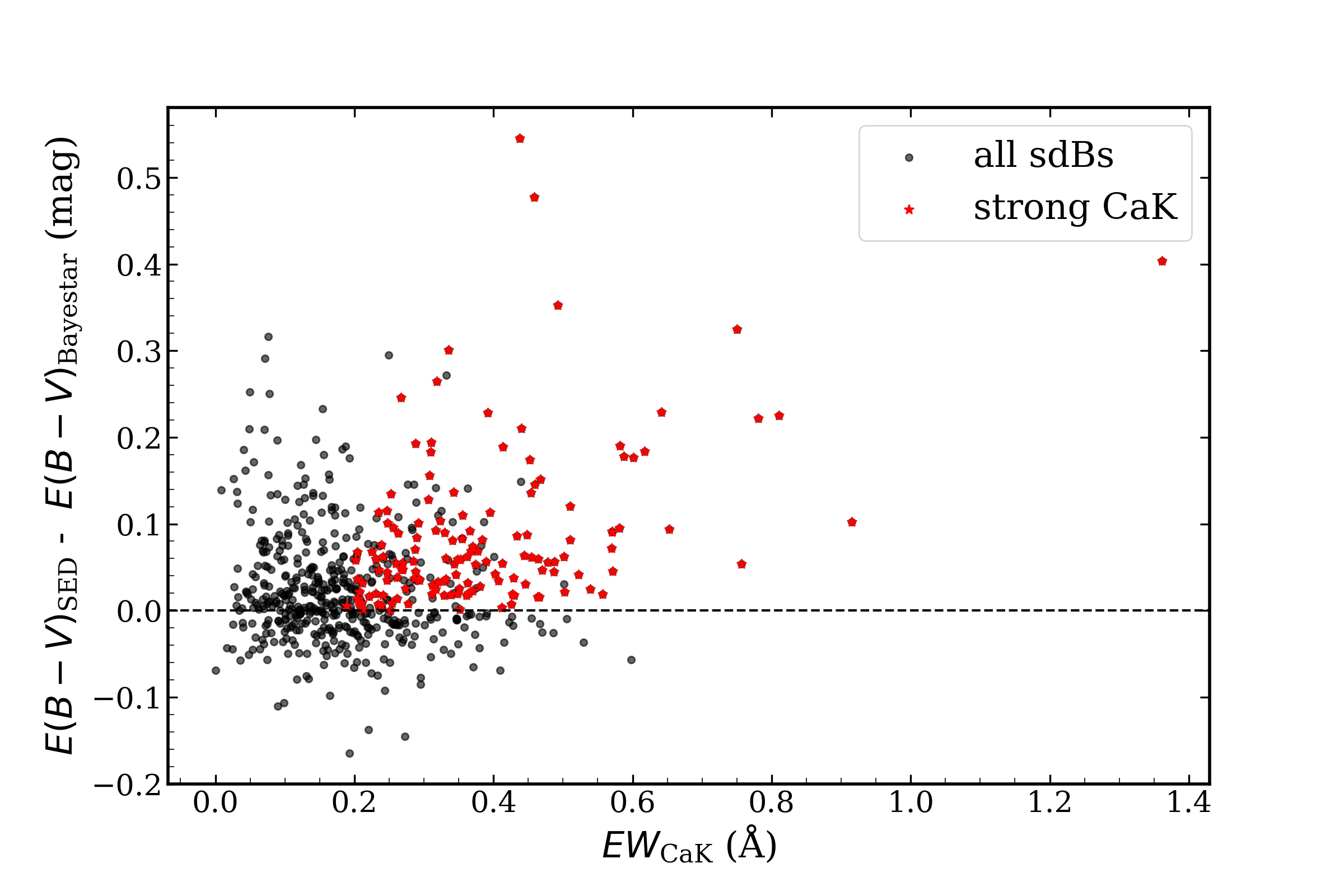}
    \caption
    {The difference in $E(B-V)$ between the values derived from SED fitting and those from the Bayestar 3D dust map. Black dots represent the 623 sdBs with detectable Ca II K absorption lines, while red stars indicate sdBs with large Ca II K EWs that lie above the linear fitting region between EW and $E(B-V)$, and for which the $E(B-V)$ from SED fitting exceeds that from the Bayestar 3D dust map.}
    \label{fig:work3_Ebv_difference}
\end{figure}

We then compared the SED-derived $E(B-V)$ values with those from the Bayestar19 dustmaps. As shown in Figure~\ref{fig:work3_Ebv_difference}, we observed that as the equivalent width of the Ca II K line ($EW_{\rm CaK}$) increases, the difference between the $E(B-V)$ values from SED fitting and Bayestar19 also increases. This suggests that higher $EW_{\rm CaK}$ values correlate with additional circumstellar material, which causes the discrepancy in reddening measurements. Thus, an increase in $EW_{\rm CaK}$ indicates an increase in circumstellar material.

For accuracy, we selected sdB stars with $E(B-V)$ values from SED fitting that exceeded those from the Bayestar19 dustmaps, as these stars are likely to have excess local dust and gas. This selection resulted in a final sample of 145 sdB stars for further analysis.

\section{Results}\label{sec:results}


\subsection{Estimation of Circumstellar Material Density} \label{subsec:density}
The column density of CSM around sdB stars can be estimated using the gas-to-dust ratio (GDR), which relates the amount of gas to dust in a given volume. By combining the observed reddening, represented as ${ E(B-V)}_{\rm SED} - {\it E(B-V)}_{\rm Bayestar}$, with the GDR, we can calculate the total column density of gas and dust in the CSM. This approach provides valuable insights into the material surrounding sdB stars. The reddening, ${\it E(B-V)}$, reflects the dust component, as it arises from the absorption and scattering of light by dust grains along the line of sight. We assume that ${E(B-V)}_{\rm Bayestar}$ accounts for the interstellar medium, while the difference ${E(B-V)}_{\rm SED} - {E(B-V)}_{\rm Bayestar}$ isolates the contribution from the circumstellar material. Using this difference, we can compute the column density of dust in the CSM. To estimate the total hydrogen column density $N_{\rm H}$ in the CSM, we relate the observed reddening to dust column density and apply the GDR to estimate the total mass of material (dust + gas) around the star.

The relationship between reddening and visual extinction $A_{\rm V}$ is given by: 
$A_{\rm V} = E(B - V) / 0.317$, 
based on the multiband relative extinction values from \citet{Wang_extinc_2019}. The visual extinction $A_{\rm V}$ is related to the dust column density, and by applying the GDR, we can estimate the total hydrogen column density. The total hydrogen column density, $N_{\rm H} = N_{\text{H\,I}} + 2N_{\text{H}_2}$, serves as a proxy for gas mass, while $A_{\rm V}$ gauges dust mass. Thus, the total hydrogen column density is: 
$N_{\rm H} = \text{GDR} \times A_{\rm V}$.

The GDR is a critical parameter for converting visual extinction into hydrogen column density. For the diffuse ISM in the Milky Way, the GDR is typically $1.87 \times 10^{21} \, \text{cm}^{-2} \, \text{mag}^{-1} $ from UV absorption and stellar reddening measurements \citep{Savage1977ApJ_ISM, Bohlin1978ApJ_ISM}. In higher column density regions like molecular clouds, the GDR may be larger, as observed by \citet{Lenz2017ApJ_ISM}, who found a GDR of 
$2.84 \times 10^{21} \, \text{cm}^{-2} \, \text{mag}^{-1}$ using H I emission line observations in the low-column-density regime.
In this study, we adopt a GDR of $(2.80^{+0.37}_{-0.34}) \times 10^{21} \, \text{cm}^{-2} \, \text{mag}^{-1}$, typical for denser gas and dust environments in molecular clouds within the Galactic plane (Li et al., in preparation). Using this GDR and the observed $A_{\rm V}$, we estimate the hydrogen column density $N_{\rm H}$. We find a mean hydrogen column density of
$N_{\rm H} = 7.6 \times 10^{20} \, \text{cm}^{-2}$, corresponding to the total column density of hydrogen nuclei in the CSM.

To estimate the surface density of the CSM, we apply a hydrogen mass fraction of $X = 0.7$, yielding a surface density of: $\Sigma \sim 1.8 \times 10^{-3} \, \text{g/cm}^2$.
This represents the total mass of gas per unit area in the circumstellar region around sdB stars. Understanding the column and surface densities of the CSM provides valuable insight into the mass and distribution of material surrounding sdB stars.

\subsection{Parameters of 727 sdB Candidates}

In Table~\ref{tab:work3_table}, we provide the comprehensive parameters for the 727 sdB candidates identified in the LAMOST low-resolution sample, which 145 sdBs are likely show strong absorption Ca II K from remnants from a common envelope phase. These parameters include critical atmospheric properties such as effective temperature ($T_{\rm eff}$), surface gravity ($\log{g}$), and helium abundance ($\log{y}$), essential for assessing their evolutionary status. We also include spatial parameters, such as distances and proper motions, which facilitate a detailed understanding of the candidates' locations within the Milky Way. The extinction coefficients are reported to account for the effects of interstellar reddening, ensuring accurate interpretations of the spectral data. A key focus of our analysis is the intensity of the Ca II K lines, which serve as indicators of circumstellar material. This information is crucial for evaluating the relationship between the sdB stars and their surrounding environments. Additionally, the measured radial velocities provide insights into the dynamics of these systems, revealing potential interactions between the sdB stars and the ejected material.

Among the total of 623 sdBs in the LAMOST-LRS dataset exhibiting Ca II K lines, our analysis successfully identifies these 145 candidates as significant objects for studying the remnants of common envelope evolution. This curated dataset will enhance our understanding of mass loss processes and environmental interactions associated with sdB stars, contributing valuable insights into their roles in stellar evolution.

\begin{table*}
    \caption{Table 1: Parameters for 727 Hot Subdwarf Stars from LAMOST-LRS DR7}
    \label{tab:work3_table}
    \setlength{\tabcolsep}{2\tabcolsep}
    \begin{tabular}{lccc}
    \hline
Num &  Label                & Units       &  Definitions \\
\hline
1   &  LAMOST              &             &  LAMOST ID\\
2   &  RA                 & deg         &  Right ascension at epoch 2000.0 (ICRS) \\
3   &  Dec                     & deg         &  Declination at epoch 2000.0 (ICRS)\\
4   & $T_{\rm eff}$          & K           & Effective temperature \\
5   & e\_$T_{\rm eff}$     & K           & Standard error in $T_{\rm eff}$ \\
6   & $\log{g}$              & dex         & Surface gravity \\
7   & e\_$\log{g}$         & dex         & Standard error in $\log{g}$ \\
8   & $\log{y}$          & dex         & Surface He abundance $y = n(\rm He)/n(\rm H)$ \\
9   & e\_$\log{y}$         & dex         & Standard error in $\log{y}$ \\
10  & $PM_{\rm RA}$         & mas $\mathrm{yr}^{-1}$          & Proper motion in right ascension \\
11  & e\_$PM_{\rm RA}$    & mas $\mathrm{yr}^{-1}$          & Standard error in $PM_{\rm RA}$ \\
12  & $PM_{\rm Dec}$        & mas $\mathrm{yr}^{-1}$          & Proper motion in declination \\
13  & e\_$PM_{\rm Dec}$    & mas $\mathrm{yr}^{-1}$      & Standard error in $PM_{\rm Dec}$ \\
14  & Dist                  & pc          & Distance \\
15  & e\_Dist           & pc          & Standard error in distance \\
16  & $U$                   & km/s        & Galactic radial velocity positive towards the Galactic centre \\
17  & e\_$U$               & km/s        & Standard error in $U$ \\
18  & $V$                   & km/s        & Galactic rotational velocity in the direction of the Galactic \\
19  & e\_$V$               & km/s        & Standard error in $V$ \\
20  & $W$                   & km/s        & Galactic velocity toward the North Galactic Pole \\
21  & e\_$W$               & km/s        & Standard error in $W$ \\
22  & $RV$                 & km/s        & Radial velocity from LAMOST-LRS spectra \\
23  & e\_$RV$              & km/s        & Standard error in $RV$ \\
24  & $RV_{\rm CaK}$        & km/s        & Radial velocity of Ca~K absorption line\\
25  & e\_$RV_{\rm CaK}$    & km/s        & Standard error in $RV_{\rm CaK}$ \\
26  & $EW_{\rm CaK}$         & {\AA}       & Equivalent width of Ca~K absorption line  \\
27  & e\_$EW_{\rm CaK}$    & {\AA}       & Standard error in $EW_{\rm CaK}$ \\
28  & $E(B-V)_{\rm 3D}$             & mag         & Reddening from 3D dustmap\\
29  & $E(B-V)_{\rm SED}$             & mag         & Reddening from SED\\
30  & e\_$E(B-V)_{\rm SED}$             & mag         & Standard error in $E(B-V)_{\rm SED}$\\
    
\hline

\end{tabular}

\end{table*}

\section{The Ubiquity of Circumbinary Material} \label{sec:ubiquity}
\subsection{Evolutionary Tracks}

\begin{figure}
    \centering
    \includegraphics[width=0.52\textwidth]{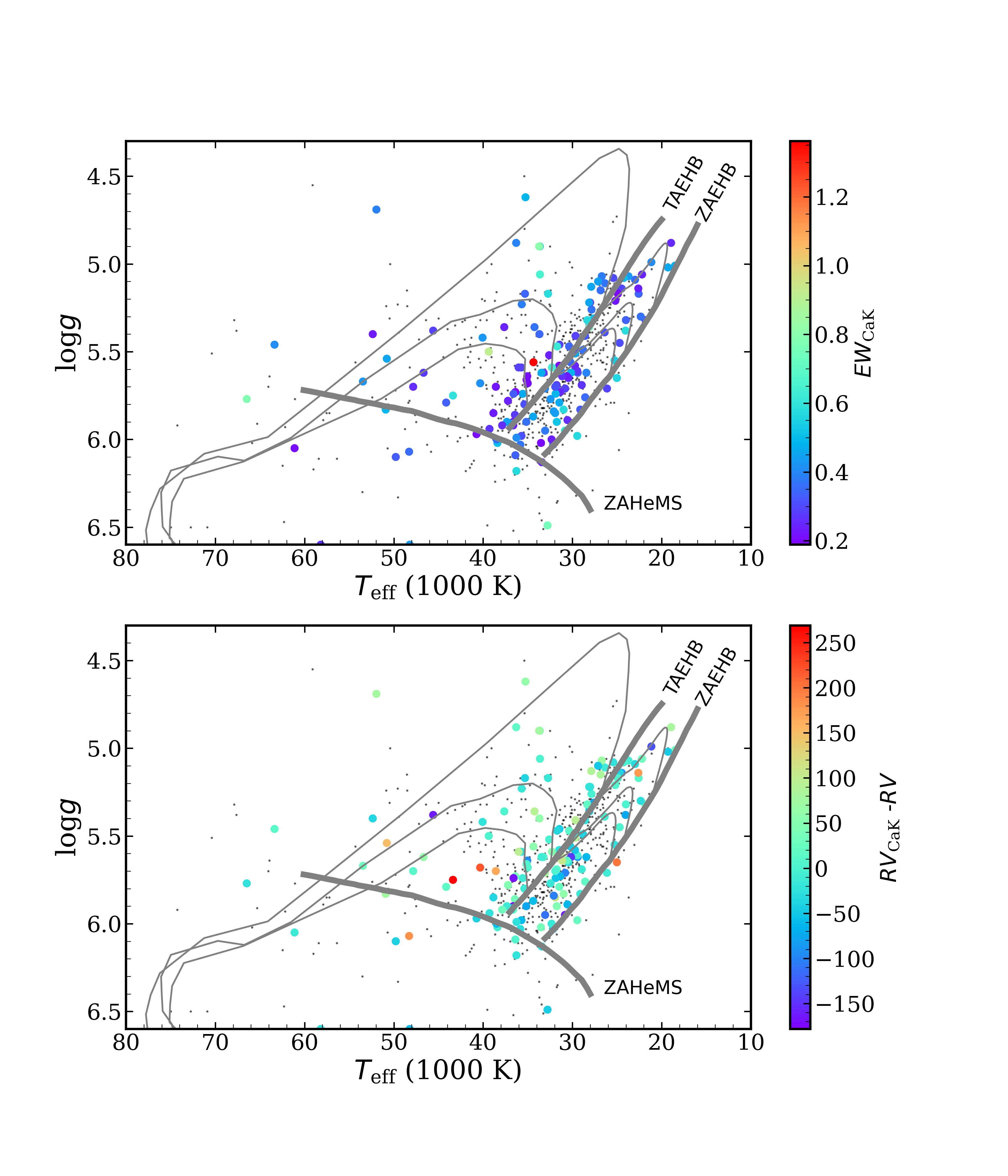}
    \caption
    {$T_{\rm eff}$–$\log{g}$ diagram for hot subdwarf stars. The thick gray solid lines indicate the zero-age extreme horizontal branch (ZAEHB) and terminal age extreme horizontal branch (TAEHB) \citep{Dorman1993ApJ_evolution_tracks}, as well as the zero-age helium main sequence (ZAHeMS) \citep{PaczynskiB1971_HeMS}. The three thin gray solid lines represent the evolutionary tracks of \citet{Dorman1993ApJ_evolution_tracks} for solar metallicity subdwarfs, corresponding to masses of 0.480, 0.473, and 0.471 ${\rm M}_{\odot}$ from top to bottom. The black dots depict the entire 727 sdB sample selected from LAMOST-LRS, while the colorful circles highlight the 145 sdBs exhibiting strong Ca II K absorption lines above the ISM level. The colors of the upper and lower panels represent the equivalent width ($EW_{\rm CaK}$) and the radial velocity difference ($RV_{\rm CaK} - RV$), respectively.
    }
    \label{fig:work3_evolution_track}
\end{figure}

We investigate the correlation between the presence of Ca II K absorption and the evolutionary stage or age of sdB stars by comparing them to theoretical evolutionary tracks. Figure~\ref{fig:work3_evolution_track} displays the distribution of hot subdwarf stars in the $T_{\rm eff}$–$\log{g}$ diagram, marking the zero-age extreme horizontal branch (ZAEHB) and terminal-age EHB (TAEHB) from \citet{Dorman1993ApJ_evolution_tracks}, along with the zero-age helium main sequence (ZAHeMS) from \citet{PaczynskiB1971_HeMS}. In this diagram, sdBs positioned closer to the ZAEHB are considered younger, while those nearer to the TAEHB are older. Three evolutionary tracks for solar metallicity subdwarfs with a core mass of 0.47 ${\rm M}_{\odot}$, as derived from \citet{Dorman1993ApJ_evolution_tracks}, are also plotted. The three thin gray solid lines represent subdwarf masses of 0.480, 0.473, and 0.471 ${\it M}_{\odot}$, respectively. The upper panel uses color to indicate varying equivalent widths ($EW_{\rm CaK}$), with redder hues corresponding to stronger $EW_{\rm CaK}$ and bluer hues indicating weaker absorption. The lower panel reflects the different radial velocity differences ($RV_{\rm CaK} - RV$) in a similar color scheme.

Our analysis reveals no significant correlation between the presence or radial velocities of Ca II K lines and the positions of sdB stars along their evolutionary tracks. This suggests that circumbinary material persists throughout the sdB stars' lifetimes, which typically span around $10^8$ years \citep{Han2002, Heber2009, Heber2016}. The enduring presence of this material across various evolutionary stages indicates that it does not dissipate rapidly after the common-envelope phase; instead, it remains gravitationally bound to the binary system.

We identified 145 sdB stars, approximately 20\% of the 727 candidates, exhibiting Ca II K absorption that exceeds ISM predictions and the reddening excess than the reddening from 3D dustmaps. This subset is likely surrounded by remnants of a common-envelope phase. If the circumbinary material were to disperse rapidly after the sdB phase — similar to planetary nebulae (PNe), which disperse over approximately $10^3$ years—the detection fraction would be significantly lower than 20\%, potentially as low as $10^{-5}$ compared to the lifespan of sdB stars \citep{Ziurys2006PNAS_PNe, Bertolami2016A&A_PNe}. Therefore, we conclude that the circumbinary material is long-lived and coexists with the sdB stars throughout their evolution.

\subsection{Radial velocity analysis}\label{sec:Radial_velocity_analysis}

Understanding the velocity offsets between the Ca II K lines and the systemic velocities of sdB binaries is crucial for determining whether these lines originate from CSM ejected during the CE phase or from absorption by the ISM. If the RV of the Ca II K lines aligns with the systemic velocity of the sdB binary, it is likely that the lines originate from the CSM. This distinction is fundamental for understanding the environments surrounding hot subdwarf stars and the evolution of post-CE binaries.

\begin{figure}
    \centering
    \includegraphics[width=0.5\textwidth]{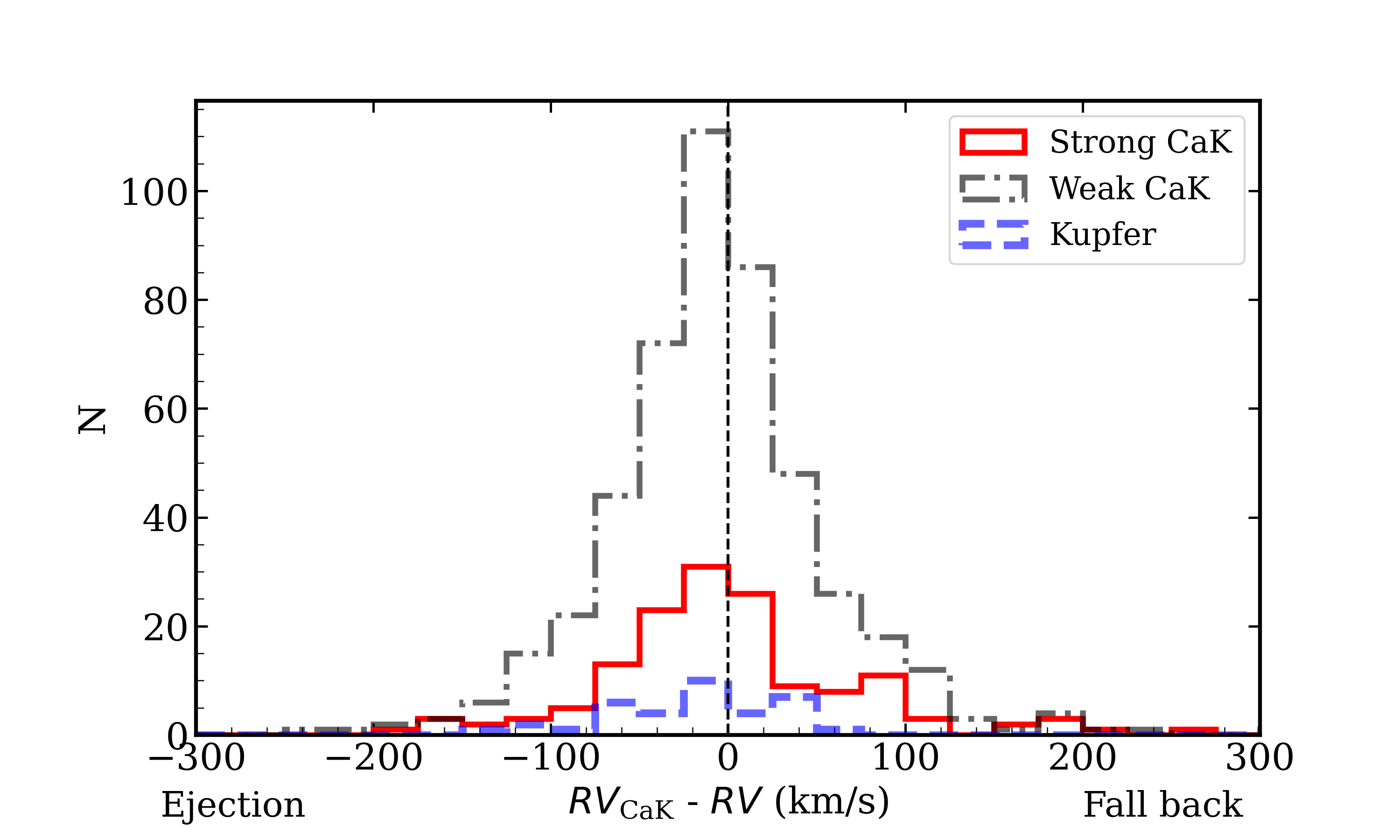}
    \caption
    {Radial velocity difference distributions. Figure compares the radial velocity differences between the Ca II K lines and the spectra across various samples: the red solid line represents sdBs with strong Ca II K lines, the black dashed-dotted line corresponds to sdBs with weak Ca II K lines, and the blue dashed line shows the sdBs star sample from \citet{Kupfer2015}. 
    }
    \label{fig:work3_RV_hist}
\end{figure}

Figure~\ref{fig:work3_RV_hist} compares RVs across different subsets of our sample, emphasizing the RV differences between the Ca II K lines and sdB spectra. The solid red line represents sdBs with strong Ca II K absorption, while the dashed-dotted black line shows those with weak absorption. Additionally, we cross-matched our entire sdB sample (including both strong and weak Ca II K absorption) with the catalogue from \citet{Kupfer2015} and identified 36 overlapping objects. For these, \citet{Kupfer2015} provided systemic velocities, allowing us to compare the RVs of the Ca II K lines with the systemic velocities. The blue dashed line illustrates the RV differences between the Ca II K lines and the systemic velocities of these 36 sdBs. 
To account for potential impacts of binary motion, we note that short-period and/or extreme mass ratio systems can show significant RV amplitudes due to orbital motion of the sdB star. However, this does not affect our results significantly, as the mean of the RV distribution is not biased by random phase observations; only the width of the distribution is inflated. By analyzing the RV distribution across the entire sample, the effects of orbital motion are minimized, as the distributions primarily reflect the systemic velocities of the binaries. Notably, the RV distributions across all three subsets—sdBs with strong Ca II K absorption, weak absorption, and systemic velocities—are remarkably similar, consistently centered around -20 km/s and 0 km/s. This uniformity strongly suggests that the Ca II K lines are a common feature of the CSM ejected during the CE phase, consistently present around sdB stars in our sample.

The observed negative RV offsets likely indicate that the circumstellar material is outflowing, likely as a result of common-envelope ejection. Stronger Ca II K absorption correlates with more pronounced outflows, reinforcing the idea that this material is dynamically moving away from the stars. Additionally, the negative offsets may be influenced by the local standard of rest velocity for Ca II K, which is typically lower than 0 km/s due to the material's trace from the circumgalactic medium \citep{Smoker2003MNRAS_ISM_cak, Smoker2015A&ACaK, Murga2015}. Since most of our sample is not located on the Galactic disk, a direct comparison with ISM radial velocities using the Galactic disk rotation curve \citep{Schoenrich2010, Schonrich2012MNRAS_Galactic_rotation} is not feasible. High-resolution infrared observations and further modelling may be needed to resolve this issue.

The small radial velocity differences between the Ca II K lines and the stellar spectra further support the ubiquity of CSM around sdB stars ejected during CE. These minor velocity offsets, well within the gravitational influence of the sdB stars and below the expected escape velocities \citep{Clayton2017}, suggest that the CSM remains bound to the system, forming a stable circumbinary envelope. This material has not dispersed into the ISM but has instead remained in orbit around the binary system, highlighting the formation of a long-lived circumbinary envelope that endures throughout the sdB star's lifetime.

To further illustrate this, we examine the case of PG0848+186, which has the highest number of observational epochs in our sample. All nine spectral observations from LAMOST-LRS have a SNR above 30 in the $g$-band. We used iSpec \citep{Blanco-Cuaresma2014a, Blanco-Cuaresma2019} in combination with TLUSTY atmospheric models \citep{Lanz2003} to generate a spectral template with $T_{\mathrm{eff}} = 25\,000$\,K and $\log g = 4.75$. The spectral resolution was set to ${\it R} \sim 1\,800$, covering the range 3\,945\,\AA\ to 5\,000\,\AA. The observed spectra were trimmed to this range, excluding the Ca II K lines but retaining the He, $H_\delta$, $H_\gamma$, and $H_\beta$ lines for velocity analysis.

\begin{figure}
    \centering
    \includegraphics[width=0.5\textwidth]{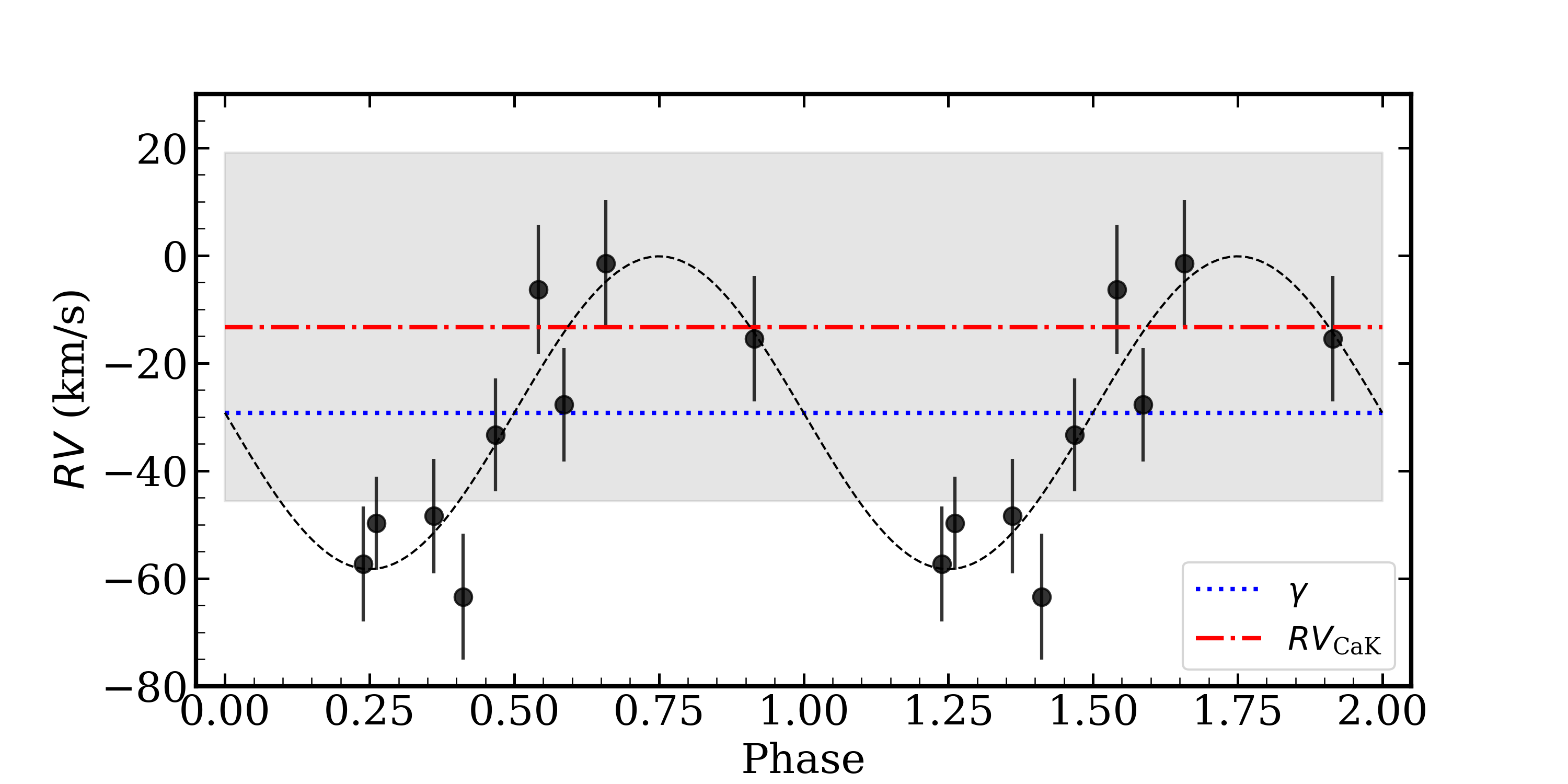}
    \caption
    {Radial velocity curve fit for PG0848+186. Black dots represent the phased radial velocities from LAMOST-LRS, and the black dashed line is the best-fit model. The orbital parameters are eccentricity $e=0$, primary star amplitude $K_1 = 29.1 \pm 4.1$\,km/s, and systemic velocity $\gamma = -29.2 \pm 3.1$\,km/s. The blue dotted line indicates the systemic velocity, while the red dot-dashed line shows the radial velocity of the Ca II K lines. The gray shaded region represents the 1$\sigma$ uncertainty of the Ca II K line velocities.
}
    \label{fig:work3_systemicRV}
\end{figure}

RVs were measured by cross-correlating the observed spectra with the template using the cross-correlation function (CCF) in iSpec. The resulting velocity shifts were converted to heliocentric RVs, which were then fitted with a sinusoidal function using the RadVel Python package \citep{Fulton2018}. Uncertainties were estimated via a MCMC analysis with 1\,000 iterations. The best-fit orbital parameters for PG0848+186 yielded an eccentricity of $e = 0$, a radial velocity semi-amplitude of $K_1 = 29.1 \pm 4.1$\,km/s, and a systemic velocity of $\gamma = -29.2 \pm 3.1$\,km/s, with an orbital period of 0.53\,d and an ephemeris zero point of 2456013.62\,d (see Figure~\ref{fig:work3_systemicRV} for the best-fit curve).
The Ca II K line measured across all nine epochs showed a constant radial velocity of $-13.3 \pm 32.4$\,km/s (red dot-dash line), which is consistent with the binary's systemic velocity ($\gamma = -29.2 \pm 3.1$\,km/s, blue dotted line) within the error margin (gray region). Importantly, the Ca II K lines exhibited no periodic shifts, suggesting a stable velocity profile across orbital phases.

These consistent RV measurements of the Ca II K lines, which closely match the systemic velocity of the binary, support the hypothesis that the lines originate from circumbinary material. The absence of significant shifts and the agreement with systemic velocities suggest that the material moves in unison with the binary. While slight redshifts in some Ca II K lines may indicate a slow inward drift of material, these observations, along with deviations from ISM predictions, confirm that the Ca II K lines trace material ejected during the common envelope phase, which remains gravitationally bound to the system. 

In conclusion, our findings suggest that CSM is a universal feature around sdB stars, ejected during the CE phase, and its dynamics align with the systemic velocities of the binaries. The consistent presence of circumbinary material around all sdBs in our sample indicates the formation of long-lived, stable circumbinary envelopes that persist throughout the sdB star's lifetime. Future high-resolution, multi-epoch spectroscopy will be essential to further constrain the dynamics of this material and its role in binary evolution.

\section{Discussions} \label{sec: Discussions}

\subsection{Mass and Radius of Circumbinary Material}

In this section, we estimate the mass and radius of the circumbinary material surrounding sdB stars by calculating densities from the $E(B-V)$ difference between the SED and 3D dust map. As previously noted, 145 sdB stars—approximately 20\% of the 727 candidates—show Ca II K absorption that exceeds ISM predictions. This subset is likely surrounded by remnants of the CE phase, and we assume the circumbinary material is long-lived, coexisting with the sdB stars. To estimate the geometric cross-section, we assume that the circumbinary material persists for at least as long as the sdB stage, with its distribution influenced by both geometry and longevity. The material’s distribution is expected to be uneven, likely due to the anisotropic nature of the circumbinary envelope. The 20\% detection rate suggests a partial solid-angle filling factor, with the material likely concentrated near the binary’s equatorial plane.

We model the circumbinary material as a disk-like envelope with random inclinations. The detection of Ca II K absorption in 145 stars suggests our line of sight intersects a wedge-shaped envelope with a biconical void. The probability of detecting this material depends on the wedge's opening angle, which dictates how much material lies along our line of sight. Based on the observed fraction of stars exhibiting Ca II K absorption, we assume the wedge angle is approximately $\pm 11.5^\circ$.

Figure~\ref{fig:work3_RV_hist} shows that the distribution of Ca II K absorption lines aligns closely with the systemic velocities of the sdB binaries, supporting the hypothesis that the circumbinary material remains gravitationally bound after common-envelope ejection. Assuming a uniform distribution of material within a wedge-like circumbinary disk with a wedge angle of $\pm 11.5^\circ$, we estimate the ejected mass using a mean column density of $1.8 \times 10^{-3} \, \mathrm{g\,cm^{-2}}$. For a typical sdB star mass of $0.47 \, {\rm M}_{\odot}$ \citep{Kupfer2015, Schaffenroth2022A&A_sdBsI}, the progenitor star mass is approximately $1 \, {\rm M}_{\odot}$ \citep{Han2002, Han2003}. Simulations of CE evolution suggest that after the fast-spiral phase, most of the CE material is ejected, with only a small fraction remaining bound to the system. Significant mass is lost as spherical ejecta, jets, or outflows \citep{Ivanova2016MNRAS, Sand2020A&A_CE, Kramer2020A&A_sdB_CEsimulations, Chamandy2020MNRAS_CE}. We assume the remnant mass after CE ejection ranges from 0.1 to 0.5 ${\rm M}_{\odot}$. Based on these assumptions, we estimate the radius of the circumbinary material to be between 31\,338\,AU and 14\,015\,AU. These estimates are derived under the assumption that the material is bound to the system and persists throughout the sdB star's lifetime. While this analysis provides valuable qualitative insights, it does not offer detailed quantitative measurements of the CSM properties. Nevertheless, the observations strongly support the conclusion that the Ca II K lines trace material ejected during the CE phase, which forms a long-lived circumbinary envelope around sdB stars.

\subsection{Bias Correction for ISM Blending}

A significant challenge in estimating Ca II K from low-resolution spectra is the blending of these lines with those produced by the ISM between the Earth and the stars. The Ca II K lines from the CSM can become indistinguishable from ISM lines, resulting in an apparent enhancement in strength. As stellar light travels through interstellar space, it experiences attenuation from absorption or scattering, particularly affecting elements like Na and Ca. The strength of ISM lines correlates with the amount of dust or gas along the line of sight, and these lines are generally narrow and sharp due to the low internal velocity dispersion of ISM gas \citep{Koo2022ApJ_Na}. 

In high-resolution spectra, ISM absorption lines can be identified and removed, especially when the radial velocity differences between the CSM and ISM gas clouds are small. However, in low-resolution spectra, such as those from LAMOST-LRS, CSM absorption lines blend with ISM lines, complicating their differentiation. This blending can lead to an overestimation of stellar elemental abundances. To evaluate the influence of ISM lines, we found that the mean equivalent width of the Ca II K line from the ISM is approximately 0.1 \AA, based on the linear relation shown in Figure \ref{fig:work3_disk_EW}. Our calculated equivalent width $EW_{\rm CaK}$ has a mean error of 0.23 \AA, significantly larger than the ISM contribution. Consequently, subtracting the ISM component from our measured Ca II K lines is unnecessary, as it does not substantially affect our results, especially since we selected samples exceeding one standard deviation above the ISM line.

When measuring the radial velocity of the Ca II K lines, ISM blending introduces potential bias. The radial velocity of the ISM is typically assumed to be zero, which can lead to an underestimation of the CSM's radial velocity due to blending effects. To quantify this impact, we generated a grid of synthetic spectra featuring Ca II K lines, assuming an ISM equivalent width of about 0.1 \AA, derived from our sample's median (see Figure~\ref{fig:work3_disk_EW}). We varied the CSM equivalent width $EW_{\rm CaK}$ between 0.1 and 0.5 \AA, reflecting typical values seen in sdB stars.

\begin{figure}
    \centering
    \setlength{\leftskip}{-15pt}
    \includegraphics[width=0.55\textwidth]{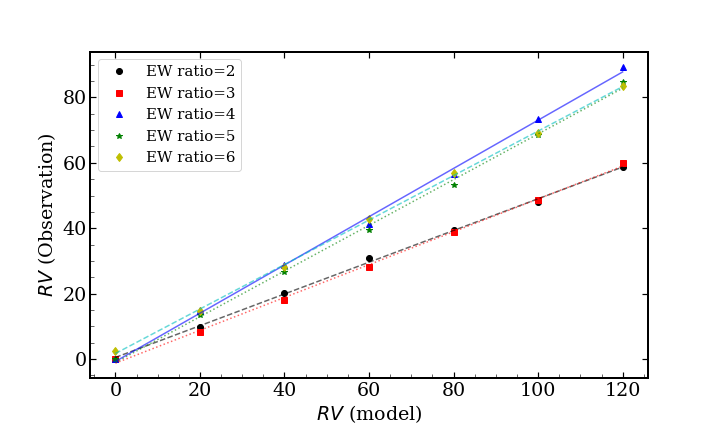}
    \caption
    {The relationship between the RV of the model for CSM and the calculated RV from the Ca II K line spectra. Black circles represent the RV correlation when the EW ratio of CSM to ISM is 2 (black dashed line). Red squares indicate the correlation at an EW ratio of 3 (red dotted line). Blue triangles correspond to an EW ratio of 4 (blue solid line). Green stars depict the relationship at an EW ratio of 5 (green dotted line), while yellow diamonds represent the correlation at an EW ratio of 6 (cyan dashed line).}
    \label{fig:work3_bias}
\end{figure}

We then adjusted the wavelength of the CSM component with radial velocities ranging from 0 to 120 km/s, as most sdBs in our sample exhibit Ca II K radial velocities between -120 and 120 km/s. The modeled spectrum combines the flux of the ISM and CSM Ca II K lines. We utilized a Gaussian fitting method to analyze the generated absorption lines, allowing us to derive observed radial velocities and equivalent widths. Our analysis revealed a relationship between the measured equivalent width $EW_{\rm CaK}$ and the expected equivalent width derived from color excess $E(B-V)$, which correlates with radial velocity shifts. The measured radial velocity can be modeled linearly with the CSM radial velocity (see Figure~\ref{fig:work3_bias}). To mitigate ISM line influence, we examined the ratio of measured $EW_{\rm CaK}$ to the expected $EW$ from $E(B-V)$. A ratio greater than 6 indicates a dominant CSM line, helping to reduce the bias from not separately fitting the ISM line.

Despite applying the correction approach, the mean error in the radial velocity measurements of the Ca II K line for the 145 sdB stars in our study is approximately 24.9 km/s, primarily due to the low spectral resolution (around 1\,800). The derived corrections for radial velocity are generally less than 20 km/s, a minor adjustment that does not significantly affect our conclusions. As such, we have chosen not to apply further corrections to the radial velocities in this study, prioritizing the robust identification of CSM contributions and emphasizing the validity of our measurements in sdB star analysis.

Moving forward, we aim to conduct observations with higher spectral resolution (above 10\,000) to better resolve the details of the Ca II K line profile. Higher resolution will allow us to better identify saturation effects and differentiate between ISM and CSM contributions based on distinct radial velocities and absorption features. Additionally, performing multiple observations of select objects will help refine our analysis by comparing radial velocities of the Ca II K line with other absorption lines, such as Na I D and K I. These advancements will enhance our understanding of line blending and spectroscopy, contributing valuable insights to the study of sdB stars and their surrounding environments.

\subsection{Na I D and K I Analysis}

The ionized phase of the ISM does not always correlate well with the neutral medium. To address this, we analyzed the profiles of the Na I and K I resonance doublets in the LAMOST spectra, aiming to distinguish circumstellar features from those of the ISM. Neutral Na I is unlikely to be observed near stars with effective temperatures exceeding $T_{\rm eff} > 20\,000$ K unless shielded by dust \citep{Lallement2003A&A_ISM}. Thus, Na I D lines can be useful for tracing ISM material, especially in cooler stars or when dust is present. \citet{Munari&Zwitter1997A&A} studied 32 stars at lower resolution, highlighting the non-linearity of the relationship between Na I D equivalent width and $E(B-V)$, following the curve of growth and discussing potential issues. At low extinctions and column densities, the equivalent width of Na I D is expected to linearly track $E(B-V)$.

To analyze the Na I D lines, we employed a double-Gaussian fitting method. The spectra were cropped to the region surrounding the doublet and fitted with a double Gaussian, which was subsequently subtracted from a linear continuum. This approach mirrors the method used for the Ca II K lines, with the key difference being the application of a double Gaussian fitting. Among the total sample of 727 sdB stars in this study, we identified 424 and 381 stars with measurable $EW$s for the Na I ${\rm D_1}$ and Na I ${\rm D_2}$ lines, respectively. Within the selected subset of 145 stars exhibiting strong Ca II K lines, we found 78 and 76 stars with available EWs for Na I ${\rm D_1}$ and Na I ${\rm D_2}$ lines, respectively. We adopted relations from the literature to link the reddening $E(B-V)$ with the strength of the Na I D absorption lines in the ISM. \citet{Poznanski2012MNRAS_NaD_dust} established a correlation between the reddening values from \citet{Schlegel_SFD1998ApJ} and the equivalent widths of ISM lines, utilizing high-resolution quasar (QSO) spectra along with low-resolution SDSS spectra. Furthermore, \citet{Murga2015} proposed a relationship between reddening and the strength of Na I D interstellar absorption based on extragalactic sources from SDSS. However, \citet{Murga2015} provided a best-fit relation only for $0 < E(B-V) < 0.08$. Therefore, we combined the $E(B-V)$ ranges from both studies, applying the \citet{Murga2015} relation for stars within $0 < E(B-V) < 0.08$:
\begin{equation}
\begin{split}
EW_{\rm NaD2} &= \frac{E(B-V)}{0.39(\pm0.09)}^{0.63(\pm0.05)}, \\
EW_{\rm NaD1} &= \frac{E(B-V)}{0.26(\pm0.02)}^{1.06(\pm0.04)},
\end{split}
\end{equation}
For stars with $E(B-V)\geq 0.08$ we adopt the relation of \cite{Poznanski2012MNRAS_NaD_dust}:
\begin{equation}
\begin{split}
EW_{\rm NaD2} &= \frac{\log_{10} (E(B-V))+1.91(\pm0.15)}{2.16}, \\
EW_{\rm NaD1} &=  \frac{\log_{10} (E(B-V))+1.76(\pm0.17)}{2.47},
\end{split}
\end{equation}
the subscript D1 in the equations refers to the Na I ${\rm D_1}$ line at 5897.56 \AA, while D2 applies to the line at 5891.58 \AA. The typical uncertainty in the estimated equivalent width ($EW$) of the Na I D lines using these relations is approximately 20\% for a given $E(B-V)$.

\begin{figure}
    \centering
    \includegraphics[width=0.5\textwidth]{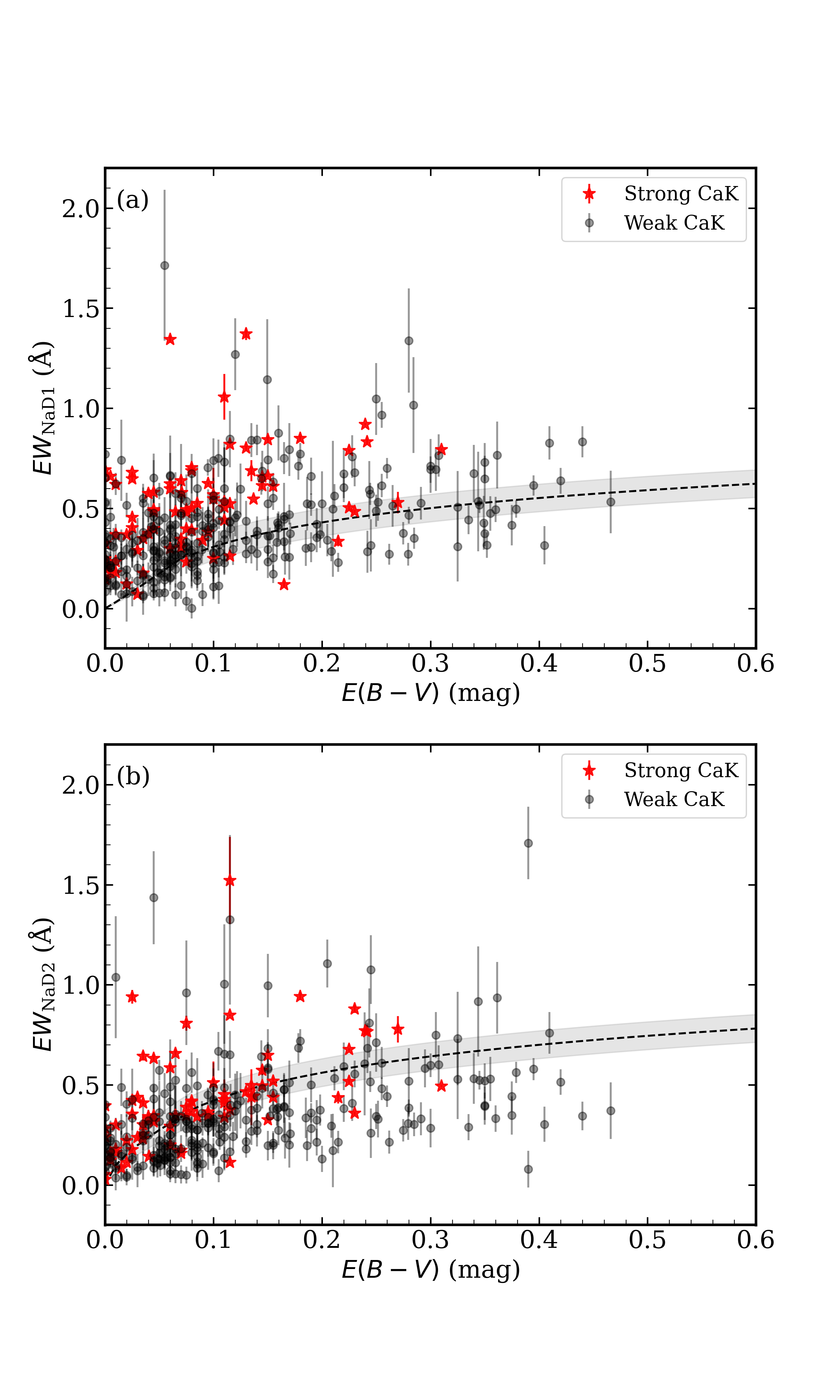}
    \caption{Relation between $EW$ of Na I D Lines and $E(B-V)$. Panel (a) shows the relationship between $E(B-V)$ and the equivalent width $EW$ of Na I ${\rm D_1}$ (5\,897.56\,{\AA}). Red stars represent stars with strong Ca II K lines, while black circles denote stars with weak Ca II K lines. The black dotted line indicates the fitted relation; for stars with $0 < E(B-V) < 0.08$, we adopted the relation from \citet{Murga2015}, and for stars with $E(B-V) \geq 0.08$, we used the relation from \citet{Poznanski2012MNRAS_NaD_dust}. The grey region indicates the typical uncertainty of the estimated $EW$ for Na I D lines, approximately 20\% for a given $E(B-V)$. Panel (b) presents the relationship between $E(B-V)$ and $EW$ of Na I ${\rm D_2}$ (5891.58\,{\AA}), with symbols and lines as in Panel (a). }
    \label{fig:work3_NaD}
\end{figure}

Figure~\ref{fig:work3_NaD}, panels (a) and (b), show that most Na I ${\rm D_1}$ and ${\rm D_2}$ equivalent widths follow the expected relations between $EW$ and $E(B-V)$ from \citet{Murga2015} and \citet{Schlegel_SFD1998ApJ}, though a few sdB stars exhibit higher $EW$s than expected. These deviations likely suggest that the material contributing to these lines may include a combination of ISM and outer CSM material.

\begin{figure}
    \centering
    \includegraphics[width=0.5\textwidth]{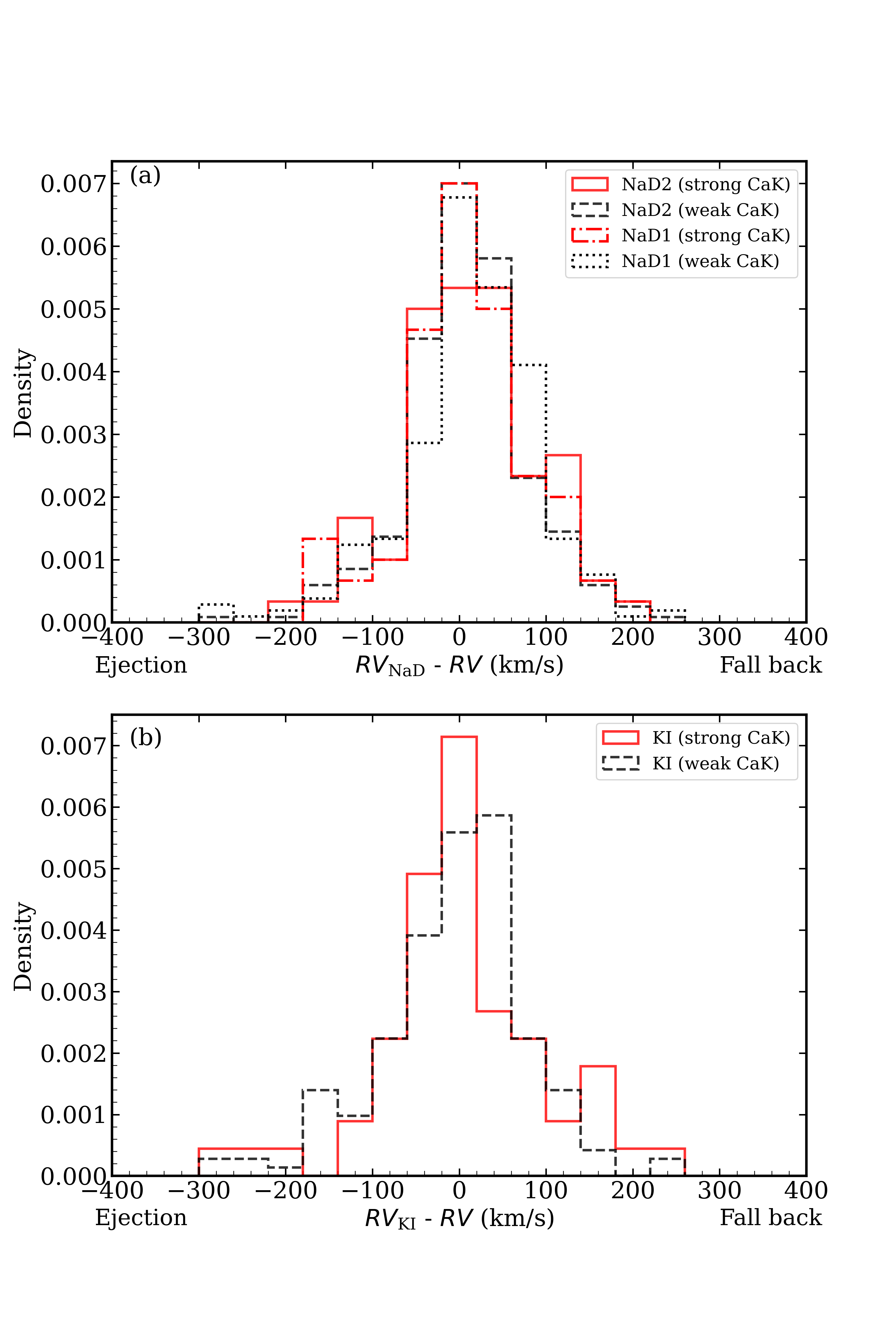}
    \caption{RV differences between Na I D / K I lines and the stellar spectra. Panel (a): Distribution of RV differences for Na I D lines. The red solid line represents Na I ${\rm D_2}$ in stars with strong Ca II K absorption, while the black dashed line corresponds to stars with weak Ca II K absorption. The red dot-dashed and black dotted lines show the distributions for Na I ${\rm D_1}$ in stars with strong and weak Ca II K absorption, respectively. Panel (b): Distribution of RV differences for K I lines. The red solid line represents stars with strong Ca II K absorption, and the black dashed line corresponds to stars with weak Ca II K absorption.}
    \label{fig:work3_KI}
\end{figure}

To further explore the origins of Na I D and K I lines, we examined their radial velocity distributions. Panel (a) of Figure~\ref{fig:work3_KI} displays the radial velocity differences between Na I D and the stellar spectra, with values clustering around 0 km/s. There is no significant difference between sdBs with strong or weak Ca II K lines, indicating that the Na I D lines primarily originate from the ISM. Only a few cases show radial velocity differences indicative of a contribution from the outer CSM.

A similar trend is observed for K I lines, as shown in panel (b) of Figure~\ref{fig:work3_KI}. From the total sample of 727 sdB stars, 311 stars have measurable $EW$s for K I lines. Within the 145 stars exhibiting strong Ca II K lines, 66 stars also show measurable $EW$s for K I. The radial velocity differences between K I and the stellar spectra are also centered around 0 km/s, with no significant distinction between sdBs with strong or weak Ca II K lines, indicating that K I lines primarily arise from the ISM as well.

These findings suggest that neutral lines, such as Na I D and K I, are mainly attributed to the ISM, with a small contribution from the outer CSM. In contrast, ionized lines like Ca II K are more likely sourced from the inner CSM. The higher excitation temperature of Ca II K compared to Na I D and K I \citep{Gray2009book} is consistent with the dynamics of a hot, ionized CSM surrounding the sdB stars. The inner CSM is expected to be hotter due to direct illumination by the central sdB star, while the outer regions, which contribute to Na I D and K I lines, are cooler and heated by the radiation from the star. This supports the conclusion that Ca II K lines trace material from the inner CSM, while Na I D and K I lines originate primarily from the ISM and outer CSM regions.

\subsection{The Nature of the Surrounding Material}

Our findings emphasize that common-envelope ejection is a widespread process in the formation of sdB stars, as evidenced by the prevalence of Ca II K absorption lines across the population (Figure~\ref{fig:work3_evolution_track}. The long-term persistence of this material, spanning up to $10^8$ years, suggests that it forms a stable envelope around the binary rather than dissipating quickly. Understanding the interactions between the binary components and the CSM provides valuable insights into the long-term stability and eventual fate of these systems. The outflowing nature of the CSM supports the scenario where binary components lose mass during the common-envelope phase, with a fraction of the ejected material remaining bound to the system (see Figure~\ref{fig:work3_RV_hist}). It is possible that the orbit gradually shrinks, with episodic ejections occurring over time. This assumption merits further discussion and the development of theoretical models for refinement.

The absence of infrared (IR) emission from the suspected circumbinary material offers key insights into its nature. One possibility is that the inner material is primarily gaseous and resides within the sublimation radius, where high temperatures prevent dust from surviving \citep{Lynn2005MNRAS_observations}. For a typical sdB star with an effective temperature of 30\,000 K and a radius of 0.2 R$_{\odot}$, the sublimation radius for silicate grains is approximately 80 R$_{\odot}$, significantly larger than the orbital separations of short-period sdB binaries. While most of the remnant mass, estimated at around 0.1 M$_{\odot}$, may be dispersed farther out, the observed column density is dominated by the small fraction of material near the star. This inner material, though limited in mass, explains the low column density and the absence of detectable IR excess in WISE observations, where faint emissions may be diluted by the instrument's large pixel size. Additionally, colder dust might emit at longer IR wavelengths beyond WISE’s sensitivity. The detection of Ca II K absorption features confirms the presence of circumbinary material, likely existing as a transient structure replenished by ongoing binary interactions or remnants of past evolution.

To further investigate the composition and distribution of circumbinary material, observations at longer IR wavelengths or in the radio range are necessary. The lack of IR emission strongly supports a gas-dominated component with minimal dust, as the distribution of metals in the gas phase does not always align with dust \citep{Murga2015}. The hot, dust-free state of the circumbinary environment, characteristic of sdB systems with stellar temperatures around 30\,000 K, produces optical spectra dominated by absorption lines, similar to B-type post-AGB stars. While current observational limits restrict the study of the most dust-obscured systems, the consistent absence of dust-driven IR emission aligns with the gaseous nature of the material. Future radio observations, complementing optical spectroscopy, could provide critical insights into the density, kinematics, and structure of the gas, confirming its gaseous nature and characterizing the remnants of ejected common envelopes. These efforts will deepen our understanding of the behavior and evolution of circumbinary material in these unique stellar environments.

\section{Conclusion} \label{sec: Conclusion}

In this study, we have explored the presence and characteristics of CSM around sdB stars, focusing on the long-lived circumbinary envelopes that likely arise from CE ejections. Our findings contribute to a better understanding of the role of CE evolution in binary star systems and the interaction between sdB stars and their surrounding environments.

Our analysis reveals that approximately 20\% of the 727 sdB candidates in the LAMOST-LRS sample exhibit Ca II K absorption lines that exceed ISM predictions, suggesting the presence of circumstellar material that has been ejected during the CE phase. We estimated the mass and radius of the circumbinary material, modeling it as a disk-like envelope with a wedge-shaped structure centered on the binary's equatorial plane. The distribution of this material is found to be anisotropic, with the detected material concentrated in a wedge with an opening angle of approximately $\pm 11.5^\circ$. Using simulations and a mean column density of $1.8 \times 10^{-3} \, \mathrm{g/cm^2}$, we estimated the radius of the circumbinary material to be between 14\,015 and 31\,338 AU, indicating that the material is bound to the system and persists over the sdB star's lifetime.

Overall, this study strengthens the hypothesis that sdB stars are surrounded by long-lived circumstellar material, likely remnants of the CE phase, and provides a detailed analysis of the spectroscopic features that trace this material. The results offer important insights into the evolution of binary star systems and the role of common-envelope ejections in shaping the environments of sdB stars. Further high-resolution observations and modeling of these systems will be crucial to refine our understanding of the CSM properties.

\section*{Acknowledgements}

We thank Uli Heber for his valuable discussions and the anonymous referee for his/her valuable comments. There are no conflicts of interest associated with this work. This work is supported by National Natural Science Foundation of China (Grant Nos. 12288102, 12125303, 12090040/3, 12403040), National Key R$\&$D Program of China (Grant No. 2021YFA1600401/3), the International Centre of Supernovae (No. 202201BC070003), Yunnan Key Laboratory (No. 202302AN360001) and the "Yunnan Revitalization Talent Support Program" Science \& Technology Champion Project (No. 202305AB350003). We also acknowledge the science research grant from the China Manned Space Project with No.CMS-CSST-2021-A10.

Guoshoujing Telescope (LAMOST) is a National Major Scientific Project built by the Chinese Academy of Sciences. Funding for the project has been provided by the National Development and Reform Commission. LAMOST is operated and managed by the National Astronomical Observatories, Chinese Academy of Sciences. 

This work has made use of data from the European Space Agency (ESA) mission Gaia (https://www.cosmos.esa.int/gaia), processed by the Gaia Data Processing and Analysis Consortium (DPAC, https://www.cosmos.esa.int/web/gaia/dpac/consortium). Funding for the DPAC has been provided by national institutions, in particular the institutions participating in the Gaia Multilateral Agreement.

This publication makes use of data products from the Two Micron All Sky Survey, which is a joint project of the University of Massachusetts and the Infrared Processing and Analysis Center/California Institute of Technology, funded by the National Aeronautics and Space Administration and the National Science Foundation.

\section*{Data Availability}

The Extended Data Table 1 of parameters for 727 hot subdwarf stars from LAMOST-LRS DR7 is available online or upon request by contacting Zhanwen Han (zhanwenhan@ynao.ac.cn) or Jiangdan Li (lijiangdan@ynao.ac.cn).



\bibliographystyle{mnras}
\bibliography{ref} 








\label{lastpage}
\end{document}